# FINAL DESIGN OF THE PRE-PRODUCTION SSR2 CRYOMODULE FOR PIP-II PROJECT AT FERMILAB*

J. Bernardini†, D. Passarelli, V. Roger, M. Parise, J. Helsper, G. V. Romanov, M. Chen,
C. Boffo, M. Kramp, F. L. Lewis, T. Nicol, B. Squires, M. Turenne
FNAL, Batavia, IL 60510, USA

## Abstract

The present contribution reports the design of the pre-production Single Spoke Resonator Type 2 Cryomodule (ppSSR2 CM), developed in the framework of the PIP-II project at Fermilab. The innovative design is based on a structure, the strongback, which supports the coldmass from the bottom, stays at room temperature during operations, and can slide longitudinally with respect to the vacuum vessel. The Fermilab style cryomodule developed for the prototype Single Spoke Resonator Type 1 (pSSR1) and the prototype High Beta 650 MHz (pHB650) cryomodules is the baseline of the current design, which paves the way for production SSR1 and SSR2 cryomodules for the PIP-II linac. The focus of this contribution is on the results of calculations and finite element analysis performed to optimize the critical components of the cryomodule: vacuum vessel, strongback, thermal shield, and magnetic shield.

## INTRODUCTION

A total of seven production SSR2 CMs will be used in the PIP-II linac [1] to accelerate H- ions from 35 MeV to 185 MeV. One ppSSR2 CM will be built during the prototyping and validation stages.

The design of the ppSSR2 CM is based on a novel concept developed at Fermilab [2, 3], the Fermilab style cryomodule, which was validated by cold testing the pSSR1 CM for PIP-II [4]. It also takes into account the standardization strategy set for the PIP-II CMs [5]. To minimize the movement of the beamline components and ancillaries during cooldown and to facilitate the assembly, the coldmass and the beamline components are based on a full-length strongback that is designed to be maintained at room temperature during operations. A High Temperature Thermal Shield (HTTS) and Low Temperature Thermal Source (LTTS), along with connections for intercepts are made available between the inner surface of the vacuum vessel and the 2 K helium to reduce radiation and conduction heat transfers. The current PIP-II beam optics design requires that each SSR2 cryomodule contains three focusing lenses and five identical SSR2 cavities, where each cavity is equipped with one high-power RF coupler, and one tuner.

Cavities and focusing lenses are supported by individual support posts, which are mounted on the strongback, located between the vacuum vessel and the HTTS. A two-phase helium pipe, running the length of the cryomodule, is connected to cavities by means of Ti-SS transition joints, and to the focusing lenses by means of thermal straps. The two-phase pipe is connected to the relief line through the top hat, and to the pumping line through the bayonets on the lateral extension of the vessel. The heat exchanger and the interfaces with 2 K relief line and pressure transducers are located on the top hat of the vacuum vessel. The interior of the vessel features an inner frame, which supports a global magnetic shield.

___________
* Work supported by Fermi Research Alliance, LLC under Contract No. DEAC02- 07CH11359 with the United States Department of Energy, Office of Science, Office of High Energy Physics.
† jbernard@fnal.gov

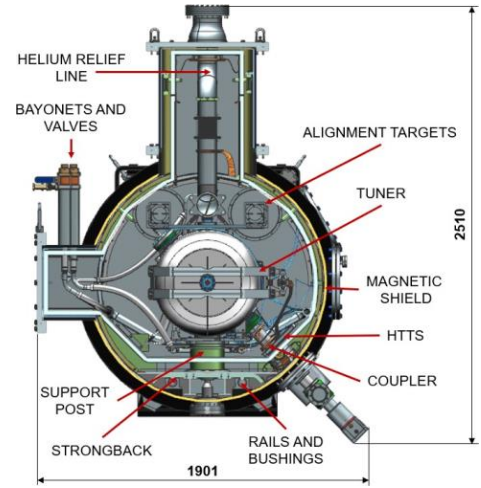

Figure 1: ppSSR2 CM transverse cross section showing the main components and subsystems (dimensions in mm).

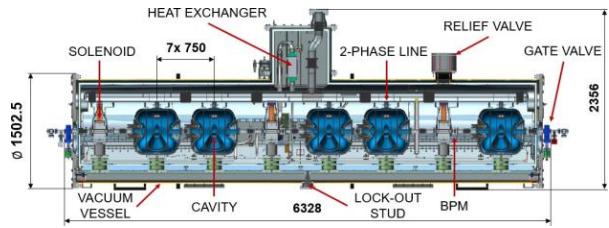

Figure 2: ppSSR2 CM longitudinal cross section showing main components and subsystems (dimensions in mm).

## MAIN CRYOMODULE COMPONENTS

### Vacuum Vessel

The vacuum vessel consists of a cylindrical shell in carbon steel (ASTM A-516) anchored to the floor with bottom supports and equipped with lugs for lifting purposes. The vessel shell is closed at the upstream and downstream side with endcaps, and it has ports for input RF power couplers, access, instrumentation, vacuum pump-out, and safety relief. The vacuum vessel provides a structural support for the strongback tray, which can slide in the vessel by means of a system of rails and bushings, as shown in Fig. 1. The strongback tray can be locked to the vessel with a central lock-out stud, as shown in Fig. 2.

Finite element analyses (FEA) were performed to evaluate the effect of the external atmospheric pressure, with vacuum inside, on the alignment of cavities and solenoids. A vessel shell thickness of 11.2 mm was considered for this analysis (the nominal thickness being 12 mm). The displacement of the beamline axis resulting from coldmass load and external pressure was probed at 16 discrete locations along the beamline's axis. Results are shown in Fig. 3. The number and position of the vessel's stiffening rings is optimized to reduce as much as possible the displacement of the beamline components. The results shown in Fig. 3 are obtained with 2 stiffening rings, which allows having a maximum displacement in the 'x' direction of 0.1 mm. The maximum displacement along 'x' with only 1 stiffening ring was calculated to be 0.25 mm. The permissible alignment errors are 1 mm RMS for the cavities, and 0.5 mm RMS for the solenoids.

Elastic stress analysis was used to validate the design against plastic collapse during CM lifting. The maximum vertical deformation of the vessel during lifting is 1.1 mm, and the highest equivalent stress occurs on the lifting bracket's weldment (each weld was modeled with a volume derated by 55%: the minimum joint efficiency specified in Table UW-12 of ASME Section VIII, Division 1 [8]). The maximum membrane plus bending stress resulting from the analysis equals 75 MPa. This value is to be compared with 1.5 times the maximum allowable stress from ASME Section II, part D, which equals 207 MPa.

Buckling analysis was performed to predict instability of the vessel under external atmospheric pressure with vacuum inside. The first mode load multiplier is 22.2, which is higher than the minimum acceptable load multiplier of 2.5.

*Strongback*

The ppSSR2 strongback consists of an Al-6061 T6 extrusion bolted to two carbon steel parallel rails and stiffened with two stainless steel I-beams. The strongback tray is designed to be pulled into the vacuum vessel by sliding the rails into open plain bearings (bushings) bolted to the vacuum vessel. A central pin prevents the strongback tray from sliding on the rails during handling, transportation, and operations. A total of 16 bushings are used in the vessel, located on the right and left of each cavity and solenoid. The number of bushings is optimized to facilitate the coldmass insertion into the vessel, and to make the strongback stiff enough to meet handling and transportation requirements.

Regarding the coldmass insertion, the maximum static deflections of the rails resulting from FEA is 0.3 mm at the most critical step of the insertion phase, as shown in Fig. 4. The bushings selected for this application have a self aligning feature that make the 0.3 mm deflection acceptable. Stresses in the strongback, rails, and stiffeners are well below the yield strengths of the materials.

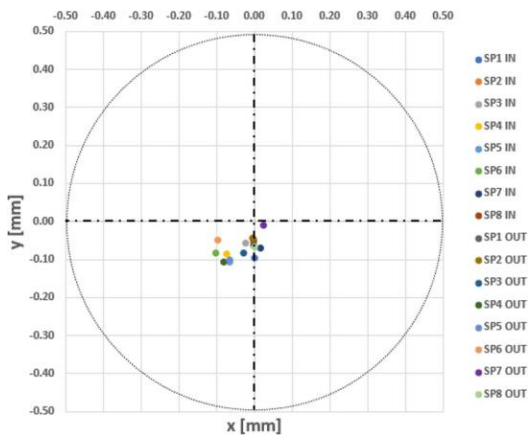

Figure 3: Cavities and solenoids transverse misalignment after pump down. The admissible misalignment is 0.5mm RMS.

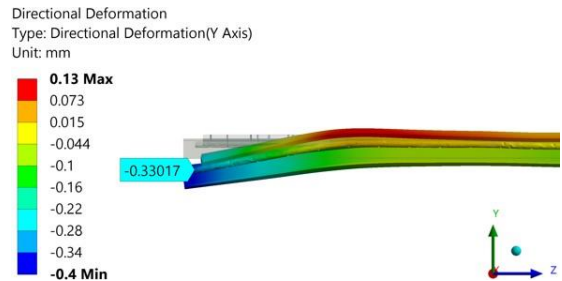

Figure 4: Strongback maximal static deflection during the insertion of the coldmass into the vessel.

Concerning handling and transportation requirements, the full CM assembly shall not have resonant frequencies below 15 Hz. The strongback's first resonant frequencies, evaluated in free-free conditions, are near 25 Hz. The 25 Hz limit is imposed by the resonant frequency of the support post, thus making the strongback stiffer would not further increase the resonant frequencies of the full CM assembly. To maintain the alignment achieved in the string assembly phase in operation, the strongback's average equilibrium temperature needs to be $T_{avg} > 283K$ with a maximum temperature differential $\Delta T < 5$ K. To make sure that the ppSSR2 strongback will remain at room temperature, one thermal strap will be installed at each coupler port location connecting the vacuum vessel to the strongback. Additionally, the bottom surface of the strongback will be painted with a vacuum compatible high emissivity coating (expected emissivity $\epsilon = 0.8$). Thermal analyses were performed to evaluate the effect of the thermal straps and the high emissivity coating on the strongback's equilibrium temperature. Analyses's results are summarized in Table 1, where $\epsilon$ is the emissivity of the aluminum surface. The analyses were performed by setting the emissivity of the internal surface of the vessel $\epsilon = 0.1$ and by imposing a 1.62 W/m$^2$ heat flux from the strongback to the HTTS.
The strongback temperature will be monitored continuously during CM testing with five temperature sensors. The movement of cavities and solenoids will be monitored continuously during the cooldown by using H-BCAMs [6,7].

Table 1: Strongback Equilibrium Temperature

| Boundary Conditions | $T_{avg}$ [K] | $\Delta T$ [K] |
|---|---|---|
| $\epsilon$=0.03 / No Straps | 285.5 | 0.8 |
| $\epsilon$=0.03 / Straps | 287.2 | 1.2 |
| $\epsilon$=0.80 / No Straps | 291.6 | 0.5 |
| $\epsilon$=0.80 / Straps | 291.8 | 0.4 |

## Thermal Shield

The HTTS aims to reduce heat loads on the LTTS and 2K volume by providing thermal intercepts and stopping radiation from room temperature components. The HTTS is composed of aluminum alloy Al 1100-H12 for the sheets and aluminum alloy Al 6061 T6 for the extrusion (the pipe carrying the helium gas). The extrusion is welded to the HTTS' sheets by means of finger welds, which are designed to reduce thermal stresses during the cooldown. The HTTS structure is supported only from the bottom: it lays on the aluminum rings shrink fitted on the support posts. The shield can slide along the longitudinal direction on seven posts and it is fully constrained on one post only. The HTTS is convection cooled by helium gas flowing in the HTTS extrusion with a nominal inlet temperature T$_{in}$ = 40 K and pressure P$_{in}$ = 13 bar. The temperature differential across the HTTS in operations shall be less than 30 K and the temperature at the interface with current leads shall be less than 65 K. A variability of ±5 K is expected in the linac on the nominal He supply temperature. Moreover, the He inlet temperature differential between the 1st and 7th cryomodule in the linac is expected to be lower than 1K. Therefore, a maximum helium inlet temperature of 46 K shall be expected for the production SSR2 CM. FE thermal analysis shows that the minimum He mass flow rate that would allow having a temperature lower than 65 K at the interface with current leads is 3.5 g/s, as shown in Fig. 5. The resulting temperature differential across the shield is 20 K.

The HTTS will be cooled down at rate of 10 K/hour. Transient thermal analysis shows that the resulting maximum temperature differential across the shield during the cooldown is 85 K, which results in a maximum membrane plus bending stress in the fillet welds of 105 MPa. The resulting stress is below the allowable limit of 130 MPa.

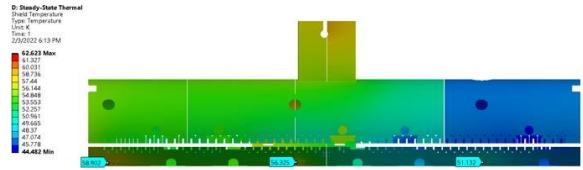

Figure 5: Thermal shield temperature distribution with a 46 K He supply temperature and 3.5 g/s mass flow rate.

## Global Magnetic Shield

The ppSSR2 CM features a global magnetic shield which is designed to attenuate the Earth's magnetic field, and the magnetic field generated by all the components internal and external to the CM, below 15 mG at the cavity surface.
The upper portion of the magnetic shield is bolted to a frame welded to the inner surface of the vacuum vessel, while the lower portion is bolted to the strongback. All the openings in the upper and lower portions will be closed as much as possible with shunts. The magnetic shield material is a 3 mm thick 80% Nickel-Iron alloy sheet, which conforms to ASTM A753-85, Type 4.
For the purpose of simulations, the magnetic permeability of the shield's material is set to $\mu$ = 40, 000, and all the components internal to the cryomodule are considered nonmagnetic. The numerical optimization of the magnetic shielding design has achieved the design goal by reducing the maximal magnetic flux density on the cavity surface down to 13.5 mG. This value further reduces to 11.7 mG when the effect of the carbon steel vacuum vessel is considered, as shown in Fig. 6.

After fitting the strongback and the magnetic shield into the vessel, measurements will be taken with a flux-gate type gaussmeter at the height of the portion of the cavities which is expected to see the highest amount of magnetic flux.

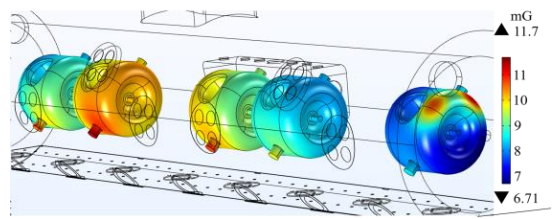

Figure 6: Magnetic flux density distribution on the surface of the cavities.

## CONCLUSION

Finite elements analysis and calculations were performed to verify that the design of the critical components of the ppSSR2 CM meets the technical requirements specifications. The design of the vacuum vessel was optimized to minimize the misalignment of cavities and solenoids during the pump down. The vacuum vessel was also proven to be safe during lifting and against buckling. Structural and thermal analyses on the strongback tray were performed to verify the feasibility of the coldmass insertion and the strongback's temperature distribution in operation, which should allow maintaining the alignment of cavities and solenoids. The HTTS design was demonstrated to meet the technical requirements given the He supply temperature and mass flow rate that can be expected in the PIP-II linac. The magnetic shield design allows having a maximal magnetic flux density of 11.7 mG on the surface of the cavities.